\documentclass[pra,print,showpacs,superscriptaddress,twocolumn]{revtex4}
\usepackage[dvips]{graphics,color}
\usepackage{amsfonts}
\usepackage{amsmath}
\usepackage{amssymb}
\usepackage{graphicx}

\input{tcilatex}
\begin{document}

\title{Nonlinear Atom-Photon Interaction Induced Population Inversion and
Inverted Quantum Phase Transition of Bose-Einstein Condensate in an Optical
Cavity}
\author{Xiuqin Zhao}
\affiliation{Institute of Theoretical Physics, Shanxi University, Taiyuan, Shanxi 030006,
China}
\affiliation{Department of Physics, Taiyuan Normal University, Taiyuan, Shanxi 030001,
China}
\author{Ni Liu}
\affiliation{School of Physics and Electronic Engineering, Shanxi University, Taiyuan,
Shanxi 030006, China}
\author{J-Q, Liang}
\email{jqliang@sxu.edu.cn}
\affiliation{Institute of Theoretical Physics, Shanxi University, Taiyuan, Shanxi 030006,
China}

\begin{abstract}
In this paper we explore the rich structure of macroscopic many-particle
quantum states for Bose-Einstein condensate in an optical cavity with the
tunable nonlinear atom-photon interaction [Nature (London) 464, 1301
(2010)]. Population inversion, bistable normal phases and the coexistence of
normal--superradiant phases are revealed by adjusting of the experimentally
realizable interaction strength and pump-laser frequency. For the negative
(effective) cavity-frequency we observe remarkably an inverted quantum phase
transition (QPT) from the superradiant to normal phases with the increase of
atom-field coupling, which is just opposite to the QPT in the normal Dicke
model. The bistable macroscopic states are derived analytically in terms of
the spin-coherent-state variational method by taking into account of both
normal and inverted pseudospin states.
\end{abstract}

\pacs{03.75.Mn, 71.15.Mb, 67.85.Pq}
\maketitle

\section{Introduction}

Quantum phase transition (QPT), which exhibits the properties of quantum
correlations, has become an exciting research field in many-body physics%
\textbf{\ }and also has important applications in quantum information
processing. The Dicke model (DM) \cite{Dic54}, which shows collective
phenomena in a light-matter system \cite{YCT07,CNL12}, is\textbf{\ }of
particular interest for the study of the fascinating QPT, since it exhibits
a second-order phase transition from a normal phase (NP) with zero average
photon-number to the superradiant phase (SP) with non-zero photons predicted
long ago \cite{WaH73,HeL73} and has broad application range \cite{ZPC08}.
The collective effects give rise to intriguing many-body phenomena such as
the existence of a coherent SP at zero temperature \cite{EmB03}. Although
the model itself is quite simple, it displays a rich variety of the unique
aspects of quantum theory and has become a paradigmatic example of
collective quantum behaviors. The DM Hamiltonian for the interaction of an
ensemble of $N$\ identical two-level atoms with single mode of the
electromagnetic field is written by \cite{EmB03,DEP07}

\begin{equation}
H_{D}=\omega _{f}a^{\dagger }a+\omega _{a}J_{z}+\frac{g}{2\sqrt{N}}\left(
a^{\dag }+a\right) \left( J_{+}+J_{-}\right) \text{,}  \label{HD}
\end{equation}%
with $\hbar =$ $1$, where $\omega _{a}$ is the frequency difference between
the two atomic levels, $\omega _{f}$ is the frequency of the cavity-field
mode, and $g$ is the atom-field dipole coupling strength. The boson
operators $a$, $a^{\dag }$ are the annihilation and creation operators for
the field, and the pseudospin\textbf{\ }$J_{i}$ ($i=z$, $\pm \ $)\textbf{\ }%
is the collective atomic operator satisfying the\textbf{\ }angular momentum
commutation relation: $\left[ J_{\pm },J_{z}\right] =\mp J_{\pm }$, $\left[
J_{+},J_{-}\right] =2J_{z}$ with the spin length $j=N/2$. The model, being a
classic problem in quantum optics, continually provides a fascinating avenue
of research in a variety of contexts. This is because the DM is a striking
example for the macroscopic many-particle quantum state (MMQS), which can be
solved rigorously. The QPT\ occurs at the critical coupling strength $g_{c}=%
\sqrt{\omega _{f}\omega _{a}}$ and\textbf{\ }the system enters a SP \cite%
{EmB03} when $g>g_{c}$.\ A significant achievement is the experimental study
of the quantum behaviors of Bose-Einstein condensates (BECs) in
ultrahigh-finesse optical cavities \cite{YCT07,BDR07}.\textbf{\ }More\textbf{%
\ }recently the time-dependent nonequilibrium experiments were performed in
an open cavity \cite{BMB11,BGB10}, which lead to the theoretical
interpretations of nonequilibrium QPT \cite{KBS10,BMS12,BER12,CLS04,NIJL13}.

It is believed that the QPT can take place only if the collective
atom-photon coupling strength is the same order of the energy separation
between the two atomic levels, which was considered as a challenging
transition-condition. In the strongly coupled regime of cavity quantum
electrodynamics (QED) this condition is shown to be accessible with the pump
laser \cite{DEP07}. For a BEC in a high-finesse optical cavity, the energy
space of two levels can be adjusted to be small enough and the QPT, namely
the superradiance transition, has been observed experimentally \cite{BDR07}.%
\textbf{\ }This is achieved by introducing two optical Raman transitions in
a four-level atomic ensemble along with the controlling of the pump laser
power \cite{BMS12}.\textbf{\ }It is shown that the theoretical model
Hamiltonian in relation with this experiment possesses a nonlinear
atom-photon interaction resulted from the dispersive shift of cavity
frequency \cite{BGB10}. For a weak nonlinear interaction, the onset of
self-organization for the ultracold atoms can be used to detect the
normal-superradiant QPT in the blue detuning of cavity frequency \cite%
{BGB10,NKS10}. This system of BEC in a\ high-finesse optical cavity has been
regarded as a promising platform to explore the exotic many-body phenomena
from atomic physics to quantum optics in a well-controlled way \cite%
{YCT07,BDR07,RicardoPuebla13,LDM08,SMorrison08,JMZhangW08,CWL08,LaM10,ZPL10,BHS09,SHB10,SND09,SND10}%
. Since the magnitude of this nonlinear interaction can arrive at the same
order as those of the detuning of cavity frequency and the collective
coupling strength, the QPT from NP to SP has been observed successfully \cite%
{BGB10}. Besides its applications in practical experiments the model of
nonlinear atom-photon interaction itself is of theoretical interest. A
natural question is whether or not the nonlinear interaction can lead to new
MMQSs compared with the standard DM of Eq. (1). It has been shown that the
nonlinear interaction, indeed, results in dynamically unstable phase \cite%
{NIJL13}. Most recently the coexistence of NP and SP was found in the
nonequilibrium QPT with time-dependent atom-field coupling \cite%
{KBS10,NIJL13} based on the numerical simulation of zero-points of the
energy functional. In order to understand the mechanism of multiphase
coexistence we revisit the generalized Dicke model of BEC-cavity experiment
\cite{BER12,NKS10} with nonlinear interaction $J_{z}a^{\dag }a$ in the whole
region of experimental parameters to reveal the bistable NPs as well as the
coexistence of NP and SP.

The Holstein-Primakoff (HP) transformation, which converts the pseudospin $%
J_{i}$ (i.e. the collective atomic operators) into a one-mode bosonic
operator, is the starting point for the most theoretical analysis of the QPT
in relation with the DM. In the thermodynamic limit ($N\rightarrow \infty $)
the DM reduces to two-mode boson Hamiltonian, the ground state of which can
be obtained in terms of variational method with the help of bosonic coherent
states \cite{EmB03,NKS10,PeV75,CLL06,AMG11,TBZ09}. We in this paper adopt
the direct product of optical and spin coherent states (SCS) \cite%
{JMR71,MVV03,AAF12} as a trial wave function firstly proposed in Ref. \cite%
{SCS} to achieve the energy functional. Based on the SCS variational method
we are able to obtain the analytical expressions of MMQSs, energy spectra,
atomic population and the photon number distribution as well. A full phase
diagram with the multistable MMQSs are presented in the whole region of
experimental parameters.

\section{ Hamiltonian for BEC in a optical cavity with nonlinear interaction
and analytic solutions}

Following the Refs. \cite{KBS10,SND10,LLM11}, we consider the system of
four-level atomic ensemble in a high-finesse optical cavity with transverse
pumping depicted in Fig. 1, where the transverse pumping laser of frequency $%
\omega _{p}$ creates a standing-wave potential and the ultracold atoms
coherently scatter pump light into the cavity mode with a position-dependent
phase.

\begin{figure}[t]
\includegraphics[width=1.5in]{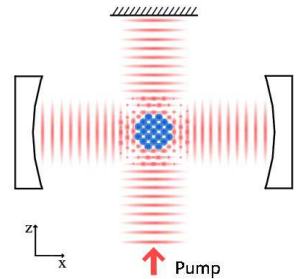}
\caption{(Color online)The experimental setup for a trapped BEC in an
optical cavity with a transverse pumping laser to control the cavity
frequency. }
\label{Fig.1}
\end{figure}

Two excited states can be eliminated adiabatically and thus we have the
effective two-level system. In an optical cavity all ultracold atoms are
assumed to couple identically with the single-mode field and the system
reduces to an extended DM given by \cite{KBS10,NKS10}.
\begin{eqnarray}
H &=&\omega a^{\dag }a+\omega _{a}J_{z}+\frac{g}{2\sqrt{N}}\left( a^{\dag
}+a\right) \left( J_{+}+J_{-}\right)  \notag \\
&&+\frac{U}{N}J_{z}a^{\dag }a\text{,}  \label{H2}
\end{eqnarray}%
where
\begin{equation}
\omega =\Delta +\beta U\text{,}
\end{equation}%
is the effective cavity frequency with
\begin{equation*}
\Delta =\omega _{f}-\omega _{p}\text{,}
\end{equation*}%
being the pump-cavity field detuning and $\beta $ is an experimental
constant \cite{KBS10,LLM11}. The nonlinear atom-photon interaction $U$
arising from the dispersive shift of cavity frequency \cite{BGB10,BER12},
can be both positive and negative values. The collective coupling strength $%
g $ is tunable in experiment by varying the pump laser power \cite{KBS10}.
The Hamiltonian Eq. (2) reduces to the standard DM Eq. (1) when the
nonlinear interaction is absent. Since the effective frequency $\omega $ can
be turned from positive to negative regions by the pump-cavity field
detuning $\Delta $ and the atom-photon interaction constant $U$, much more
rich phases arise compared with the ordinary DM.

We reinvestigate the MMQSs and related QPT based on the SCS variational
method with advantage that both the normal ($\Downarrow $) and inverted ($%
\Uparrow $) pseudospin states revealed in the dynamic study \cite{BMS12} can
be taken into account in order to see the multiple steady states observed in
the nonequilibrium QPT \cite{BMS12,BER12,NIJL13}. Moreover, an energy
functional with one-parameter only can be obtained and thus the stability of
MMQSs is justified rigorously.

\subsection{Spin coherent-state variational method}

We begin with the average of Hamiltonian Eq. (\ref{H2}) in the optical
coherent state $\left\vert \alpha \right\rangle $

\begin{eqnarray}
H_{sp}\left( \alpha \right) &=&\left\langle \alpha \right\vert H\left\vert
\alpha \right\rangle =\omega \gamma ^{2}+\omega _{a}J_{z}+\frac{U}{N}%
J_{z}\gamma ^{2}  \notag \\
&&+\frac{g\gamma \cos \eta }{\sqrt{N}}\left( J_{+}+J_{-}\right) \text{,}
\end{eqnarray}%
where $\alpha $ is the complex eigenvalue of photon annihilation operator $a$
such that $a\left\vert \alpha \right\rangle =\alpha \left\vert \alpha
\right\rangle $ and can be generally expressed as
\begin{equation*}
\alpha =\gamma e^{i\eta }\text{.}
\end{equation*}%
The effective spin Hamiltonian $H_{sp}\left( \alpha \right) $ possesses two
macroscopic eigenstates namely the SCSs $|\mathbf{n}_{\mp }>$ of south and
north pole gauges respectively, which correspond to the normal ($\Downarrow $%
) and inverted ($\Uparrow $) pseudospin states in the dynamics of
nonequilibrium DM \cite{BMS12}. The SCSs can be generated from the maximum
Dicke states $|j,\pm j>$ ( $J_{z}$ $|j,\pm j>=\pm j|j,\pm j>$) with the SCS
transformation \cite{CLS04,LLM96}, such that
\begin{equation*}
\left\vert \mathbf{n}_{\pm }\right\rangle =R(\mathbf{n})|j,\pm j>\text{,}
\end{equation*}%
where the unitary operator is explicitly given by
\begin{equation}
R(\mathbf{n})=e^{\frac{\theta }{2}\left( J_{+}e^{i\phi }-J_{-}e^{-i\phi
}\right) }\text{.}
\end{equation}%
As a matter of fact the SCSs of north and south pole gauges are actually the
eigenstates of the spin projection operator $\mathbf{J}\cdot \mathbf{n}%
\left\vert \mathbf{n}_{\pm }\right\rangle =\pm j\left\vert \mathbf{n}_{\pm
}\right\rangle $, where\textbf{\ }$\mathbf{n}=\left( \sin \theta \cos \phi
,\sin \theta \sin \phi ,\cos \theta \right) $ is the unit vector with the
directional angles $\theta $ and $\phi $. In the SCS the spin operators
satisfy the minimum uncertainty relation, for example, $\Delta J_{+}\Delta
J_{-}=<J_{z}>/2$ and therefore the SCSs $|\mathbf{n}_{\pm }>$ are called the
macroscopic quantum states. The two macroscopic eigenstates of north and the
south pole gauges are orthogonal i.e. $<\mathbf{n}_{+}|\mathbf{n}_{-}>=0$.
It is the key point to take into account of the both macroscopic eigenstates
$|\mathbf{n}_{\pm }>$ for revealing the multistable phases. Using the
unitary transformations $R(\mathbf{n})$ for the spin operators $J_{z}$, $%
J_{+}$, $J_{-}$
\begin{equation*}
\begin{array}{l}
\widetilde{J_{z}}=J_{z}\cos \theta +\frac{1}{2}\sin \theta (J_{+}e^{-i\phi
}+J_{-}e^{i\phi })\text{,} \\
\widetilde{J_{+}}=J_{+}\cos ^{2}\frac{\theta }{2}-J_{-}e^{2i\phi }\sin ^{2}%
\frac{\theta }{2}-J_{z}e^{i\phi }\sin \theta \text{,} \\
\widetilde{J_{-}}=J_{-}\cos ^{2}\frac{\theta }{2}-J_{+}e^{-2i\phi }\sin ^{2}%
\frac{\theta }{2}-J_{z}e^{-i\phi }\sin \theta \text{,}%
\end{array}%
\end{equation*}%
where $\widetilde{J_{z}}=R^{\dag }(\mathbf{n})J_{z}R(\mathbf{n})$ etc., the
effective spin Hamiltonian $H_{sp}\left( \alpha \right) $ is diagonalized
under the conditions:

\begin{equation}
\begin{array}{l}
\Phi \sin \theta e^{i\phi }+\frac{g\gamma }{\sqrt{N}}(\cos ^{2}\frac{\theta
}{2}-e^{2i\phi }\sin ^{2}\frac{\theta }{2})\cos \eta =0\text{,} \\
\Phi \sin \theta e^{-i\phi }+\frac{g\gamma }{\sqrt{N}}(\cos ^{2}\frac{\theta
}{2}-e^{-2i\phi }\sin ^{2}\frac{\theta }{2})\cos \eta =0\text{,}%
\end{array}%
\end{equation}%
with $\Phi =\frac{\omega _{a}}{2}+\frac{U}{2N}\gamma ^{2}$. From the above
conditions Eq. (6) the angle parameters $\theta $, $\phi $ can be determined
in principle. Thus we obtain the energy functional for the normal ($%
\Downarrow $) and inverted ($\Uparrow $) states respectively
\begin{equation}
E_{\mp }\left( \alpha \right) =<\mathbf{n}_{\mp }|H_{sp}|\mathbf{n}_{\mp
}>=\omega \gamma ^{2}\mp \frac{N}{2}A(\alpha ,\theta ,\phi )\text{,}
\label{7}
\end{equation}%
where%
\begin{equation*}
A(\alpha ,\theta ,\phi )=\left( \omega _{a}+\frac{U}{2N}\gamma ^{2}\right)
\cos \theta -\frac{2g}{\sqrt{N}}\sin \theta \cos \eta \cos \phi \text{.}
\end{equation*}%
The desired MMQSs%
\begin{equation*}
|\psi _{\mp }>=|\alpha >\left\vert \mathbf{n}_{\mp }\right\rangle
\end{equation*}
and corresponding energies are found as local minima of the energy
functional $E_{\mp }\left( \alpha \right) $.

\subsection{Energy functions, atomic population and mean photon numbers}

Using the diagonalization conditions Eq. (6) to eliminate the parameters $%
\cos \eta $\textbf{\ }and\textbf{\ }$\cos \phi $\textbf{, }we obtain%
\begin{equation*}
\cos \theta =\frac{\left( \omega _{a}+\frac{U}{N}\gamma ^{2}\right) }{%
A(\gamma )}\text{,}
\end{equation*}%
where the function $A(\alpha ,\theta ,\phi )$ in the energy functional Eq.
(7) becomes a one-parameter function only
\begin{equation*}
A(\gamma )=\sqrt{\omega _{a}^{2}+\frac{2}{N}(\omega _{a}U+2g^{2})\gamma ^{2}+%
\frac{U^{2}}{N^{2}}\gamma ^{4}}\text{.}
\end{equation*}%
The energy functional Eq. (7) reduces after a tedious algebra to that of one
variable i.e. $\gamma =|\alpha |$
\begin{equation}
E_{\mp }(\gamma )=\omega \gamma ^{2}\mp \frac{N}{2}A(\gamma )\text{,}
\end{equation}%
which is a key point of our approach. The MMQS solutions are found from the
usual extremum condition of the energy function
\begin{equation}
\frac{\partial E_{\mp }}{\partial \gamma }=\gamma \left[ 2\omega \mp \frac{%
(\omega _{a}+\frac{U}{N}\gamma ^{2})U+2g^{2}}{A(\gamma )}\right] =0\text{.}
\end{equation}%
The extremum condition Eq. (9) possesses always a zero photon-number
solution $\gamma =0$, which gives rise to the NP ($\gamma _{n\mp }=0$) only
if it is stable with the positive second-order derivative, namely
\begin{equation}
\frac{\partial ^{2}E_{\mp }(\gamma _{n\mp }=0)}{\partial \gamma ^{2}}%
=2\omega \mp \left( U+\frac{2g^{2}}{\omega _{a}}\right) >0\text{.}
\label{10}
\end{equation}%
We denote the NP states $\gamma _{n\mp }=0$ by $N_{\mp }$ respectively. The
non-zero photon-number solutions are found as
\begin{equation}
\gamma _{s\mp }^{2}=\frac{N}{U^{2}}\left[ -(2g^{2}+U\omega _{a})\pm \frac{%
4g|\omega |}{\varsigma }\sqrt{\xi \varsigma }\right] \text{,}  \label{AA12}
\end{equation}%
where
\begin{equation*}
\xi =g^{2}+U\omega _{a}\text{,}\quad \varsigma =4\omega ^{2}-U^{2}\text{.}
\end{equation*}%
The second-order derivative in the solutions $\gamma _{s\mp }^{2}$ is also
derived analytically%
\begin{equation}
\frac{\partial ^{2}E_{\mp }(\gamma _{s\mp }^{2})}{N\partial \gamma ^{2}}=\pm
\varsigma \sqrt{\frac{\varsigma }{\xi }}\frac{\gamma _{s\mp }^{2}}{g}\text{.}
\end{equation}%
The SPs denoted by $S_{\mp }$ are realized from both positive photon-number
Eq. (\ref{AA12}) and the second-order derivative Eq. (12). Substituting the
nonzero photon solutions Eq. (\ref{AA12}) back to the extremum condition Eq.
(9), it is easy to find the necessary conditions
\begin{equation}
\quad \omega >0\text{,}\emph{\quad }\varsigma >0\text{,}
\end{equation}%
and
\begin{equation}
\quad \omega <0\text{,}\quad \varsigma <0\text{,}  \label{14}
\end{equation}%
to be fulfilled respectively for the normal state ($\Downarrow $) solution $%
\gamma _{s-}^{2}$, and the inverted state ($\Uparrow $) $\gamma _{s+}^{2}$.
The normal state solution $\gamma _{s-}^{2}$ reduces exactly to that of DM
at the limit $U\rightarrow 0$. While the solution $\gamma _{s+}^{2}$ for the
inverted state ($\Uparrow $) is only possible when\textbf{\ }$U\lneqq 0$ and
thus is induced by the nonlinear interaction only. The critical lines can be
fixed from the equations
\begin{equation*}
\frac{\partial E_{\mp }(\gamma =0)}{\partial \gamma }=0\text{,}
\end{equation*}%
which lead to the phase boundaries
\begin{equation}
g_{c\mp }=\sqrt{\left( \pm \omega -\frac{U}{2}\right) \omega _{a}}\text{.}
\label{15}
\end{equation}%
When $U=0$\textbf{, }the critical pint for the normal state ($\Downarrow $)
approaches the typical value of DM that $g_{c-}=\sqrt{\omega \omega _{a}}$,
which is valid only for the positive effective frequency $\omega >0$.
Substituting the photon number obtained in Eq. (\ref{AA12}) into the energy
function Eq. (8) we achieve the mean energies of per atom for the SP

\begin{eqnarray}
\varepsilon _{s\mp } &=&\frac{E_{s\mp }}{N}=\frac{1}{U^{2}}\left[
-(2g^{2}+U\omega _{a})\pm 4g\omega \sqrt{\frac{\xi }{\varsigma }}\right]
\omega  \notag \\
&&\mp g\sqrt{\frac{\xi }{\varsigma }}\text{,}
\end{eqnarray}%
which become at the critical lines $g_{c\mp }$
\begin{equation*}
\varepsilon _{s\mp }(\gamma _{s\mp }^{2}=0)=\mp \frac{\omega _{a}}{2}\text{,}
\end{equation*}%
for the normal ($\Downarrow $) and inverted ($\Uparrow $) states
respectively. The mean photon number of SP in the wave functions $|\psi
_{\mp }\rangle $ is obviously
\begin{equation}
n_{p\mp }=\frac{\left\langle \psi _{\mp }|a^{\dag }a|\psi _{\mp
}\right\rangle }{N}=\gamma _{s\mp }^{2}\text{,}  \label{AU17}
\end{equation}%
while the atomic population imbalance becomes
\begin{equation}
\Delta n_{a\mp }=\frac{\left\langle \psi _{\mp }|J_{z}|\psi _{\mp
}\right\rangle }{N}=\frac{1}{U}\left[ -\left\vert \omega \right\vert \pm
\frac{g}{2}\sqrt{\frac{\varsigma }{\xi }}\right] \text{,}  \label{JU18}
\end{equation}%
which reduces to the well known values
\begin{equation*}
\Delta n_{a\mp }(\gamma _{s\mp }^{2}=0)=\mp \frac{1}{2}\text{,}
\end{equation*}%
at the critical lines $g_{c\mp }$. A full atomic population inversion i.e. $%
\Delta n_{a}=1/2$ is found in the inverted state ($\Uparrow $). It may be
worth while to remark that all these formula for the normal state ($%
\Downarrow $) reduce to those in the standard DM when $U\rightarrow 0$ and $%
\omega =\omega _{f}$ in Hamiltonian (\ref{H2}), where the critical point of
QPT is the well known form $g_{c}=\sqrt{\omega _{f}\omega _{a}}$.

\section{Bistability and atomic population inversion}

We have presented in the previous section the general formula and now show
the particular MMQSs in the experimental parameter-values. Following the
experiment \cite{BGB10} the nonlinear interaction value spreads in a wide
region $U\in $ $(-80\sim 80)$ $MHz$ (the unit of energy and frequency is $%
MHz $ throughout the paper) with the atomic frequency $\omega _{a}=1$ and
the collective atom-field coupling strength $g$ $\in $ $(0\sim 10)$. We
first consider the blue detuning of the pump-cavity field $\Delta =-20$.
With the experimental constant \cite{LLM11} $\beta =7/6$, the\textbf{\ }%
effective frequency becomes
\begin{equation}
\omega =(-20+\frac{7U}{6})\text{.}
\end{equation}%
The phase boundary for the normal state ($\Downarrow $) is found from Eq.
(15)%
\begin{equation}
g_{c-}=\sqrt{2(-10+\frac{U}{3})}\text{,}
\end{equation}%
which increases with the nonlinear interaction $U$ in exact agreement with
the previous observation \cite{LLM11}. Below $g_{c-}$ we have only the phase
$N_{-}$ shown in Fig. 2. The SP of $S_{-}$ exists in the region of $g>g_{c-}$
and $U>30$ (the value can be also evaluated from the necessary condition Eq.
(13) i.e. $\varsigma >0$). The phase boundary for the inverted state ($%
\Uparrow $) obtained from Eq. (15) is%
\begin{equation}
g_{c+}=\sqrt{5(4-\frac{U}{3})}\text{,}
\end{equation}%
which increases with the decreasing $U$ as shown in Fig. 2. The NP of $N_{+}$
for the inverted state ($\Uparrow $) always exists above the boundary curve $%
g_{c+}$, so that we have the bistable NPs denoted by NP$_{bi}$($N_{-}$, $%
N_{+}$) in Fig. 2, in which $N_{-}$ with the lower energy is the ground
state. Correspondingly the notation SP$_{co}$($S_{-}$, $N_{+}$) means the SP
of $S_{-}$ coexisting with $N_{+}$ since $S_{-}$ is the lower energy state.%
\emph{\ }The bistable MMQSs observed in this paper agree with the dynamic
study of nonequilibrium QPT \cite{BMS12,NIJL13}.

\begin{figure}[t]
\includegraphics[width=2.0in]{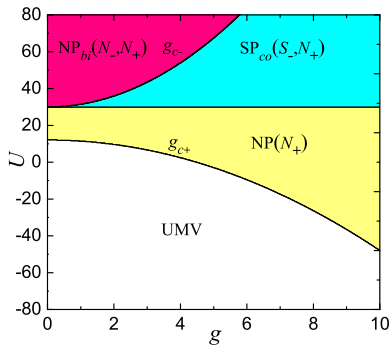}
\caption{(Color online) Phase diagram in blue detuning with $\Delta =-20$. NP%
$_{bi}$($N_{-}$,$N_{+}$) indicates the bistable NPs and SP$_{co}$($S_{-}$,$%
N_{+}$) means the SP of $S_{-}$ coexisting with $N_{+}$. The single NP of $%
N_{+}$ with full population inversion is located between the line of $U=30$
and phase boundary $g_{c+}$. The unstable macroscopic vacuum (UMV) is found
under the curve $g_{c+}$. }
\label{Fig.2}
\end{figure}
In the area (yellow) between the line of $U=30$ and curve $g_{c+}$ we have
only the single NP of $N_{+}$ with full population inversion, which induced
by the nonlinear interaction is a new observation. The SP of $S_{+}$ for the
inverted state ($\Uparrow $) does not exist, since the necessary condition
Eq. (14) can not be fulfilled in the blue detuning. Below the phase boundary
$g_{c+}$ the zero photon-number solution ($\gamma =0$) is unstable with
negative second-order derivative Eq. (10) and we call it the unstable
macroscopic vacuum (UMV). The QPT in DM is characterized by the average
photon number $n_{p}$ (or $\gamma $), which serves as an order-parameter,
with $n_{p}>0$ for the SP and $n_{p}=0$ in the NP. In Fig. 3 we present the
average photon number $n_{p}$ (a), atomic population imbalance $\Delta n_{a}$
(b) between the two atom-levels, and the average energy $\varepsilon $ (c)
as a function of the atom-field coupling strength $g$, where (black) solid
lines are for the normal state ($\Downarrow $) and (red) dot and dash lines
for the inverted state ($\Uparrow $) throughout the paper.

\ \textbf{\ }
\begin{figure}[t]
\includegraphics[width=2.0in]{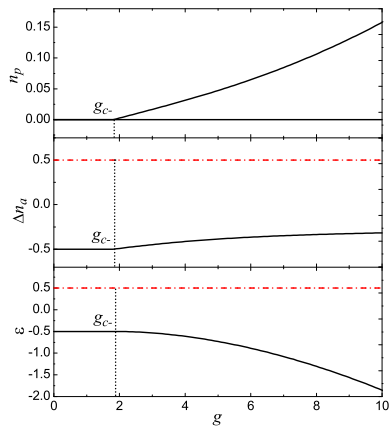}
\caption{(Color online) The variations of average photon number $n_{p}$ (a),
population imbalance $\Delta n_{a}$ (b) and average energy $\protect%
\varepsilon $ (c) with respect to the coupling constant $g$ for $U=35$ in
the blue detuning of $\Delta =-20$, with (black) solid lines for the the
normal state ($\Downarrow $), (red) dot and dash lines for the inverted
state ($\Uparrow $). The critical point of QPT from NP of $N_{-}$ to the SP
of $S_{-}$ is $g_{c-}=\protect\sqrt{10/3}$ in the given $U$ value.}
\label{Fig.3}
\end{figure}
The QPT from NP of $N_{-}$ to SP of $S_{-}$ is the standard DM type, while
the nonlinear interaction $U$ only shifts the critical point $g_{c-}$ toward
the higher value direction of the atom-field coupling $g$ \cite{LLM11}. For
the given value of $U=35$ in Fig. 3 the critical point of QPT can be
evaluated precisely for the Eq. (20) that $g_{c-}=\sqrt{10/3}$. The NP of $%
N_{+}$ remains not changed through the critical point $g_{c-}$, so that it
is the bistable NP$_{bi}$($N_{-}$, $N_{+}$) below $g_{c-}$ while the
coexistence phase of SP$_{co}$($S_{-}$, $N_{+}$).

\section{Inverted quantum phase transition}

It is an interesting aspect of the nonlinear interaction to see whether or
not the SP of $S_{+}$ for the inverted state can be realized in the
practical experiment. To this end we now turn to the red detuning ($\omega
_{p}$ $<\omega _{f}$) with, for example, $\Delta =20$. The phase boundary
lines are respectively found from Eq. (15) as
\begin{equation}
\quad g_{c-}=\sqrt{2(10+\frac{U}{3})}\text{,}
\end{equation}%
and
\begin{equation}
g_{c+}=\sqrt{-5(4+\frac{U}{3})}\text{,}
\end{equation}%
showing in Fig. 4. Two lines cross at the point $g_{c-}=g_{c+}$ ($U=-120/7$%
), which appears as a critical point of multiple phases. The SP of $S_{-}$
with $\gamma _{s-}^{2}>0$ coexists with the NP of $N_{+}$ in the region of $%
g>g_{c-}$ and $U>-12$ determined from the necessary condition Eq. (13) (cyan
area in Fig. 4). The the bistable NP$_{bi}$($N_{-}$, $N_{+}$) is located in
the area for $g\leq g_{c-}$ (pink area in Fig. 4). The single NP of $N_{+}$
with full atomic population inversion is bounded by the line of $U=-12$ and
the critical curves $g_{c-}$, $g_{c+}$ (the yellow area in Fig. 4). The
label (1) indicates a small region of the single NP of $N_{-}$. The SP of $%
S_{+}$ with the stable solution of $\gamma _{s+}^{2}>0$ is found in the
region of $U>-30$ and $U<-17$ when $g<g_{c+}$, which is below the critical
boundary line $g_{c+}$ just opposite to the normal state ($\Downarrow $)
case. The notation NP$_{co}$($N_{-}$, $S_{+}$) marked by label (2) in Fig.
(4) means the NP of $N_{-}$ coexisting with $S_{+}$ since $N_{-}$ is the
lower energy state.

\begin{figure}[t]
\includegraphics[width=2.0in]{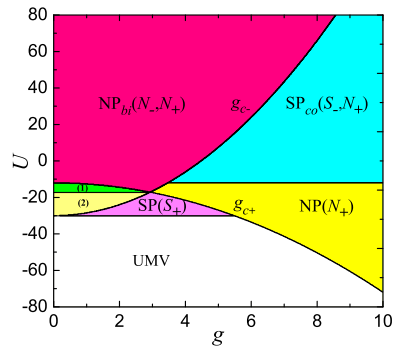}
\caption{(Color online) Phase diagram in red detuning with $\Delta =20$. The
SP of $S_{+}$ for the inverted state ($\Uparrow $) appears when the
effective frequency becomes negative $\protect\omega <0$. (1) indicates the
single NP of $N_{-}$, (2) is for NP$_{co}$($N_{-},S_{+}$).}
\label{Fig.4}
\end{figure}

The QPT from NP to SP for the standard DM is along the increasing direction
of coupling constant $g$, while the QPT in the inverted state ($\Uparrow $)
would be in the opposite direction seen from the phase diagram in Fig. 4,
where the SP of $S_{+}$ is located on the left-hand side of the critical
line $g_{c+}$. Fig. 5 displays the curves of average photon number $n_{p}$%
\textbf{\ }(a), atomic population imbalance\textbf{\ }$\Delta n_{a}$ (b),
and the energy $\varepsilon $ (c) for $U=-20$. The critical point $g_{c-}=2%
\sqrt{5/3}$ separates the coexistence phase of NP$_{co}$($N_{-}$, $S_{+}$)
and the single SP of $S_{+}$. Then with the increase of $g$ the SP of $S_{+}$
transits to NP of $N_{+}$ at the critical point $g_{c+}=2\sqrt{10/3}$. The
QPT from the SP of $S_{+}$ to the NP of $N_{+}$ is just in the inverted
direction compared with standard DM. One should not be so surprised by this
inverted QPT, since the effective frequency is negative $\omega <0$ in the
region, where the the SP of $S_{+}$ exists. By adjusting the nonlinear
constant, for example $U=-14$, the QPT among NPs of different types can be
also realized, which is displayed in Fig. 6. The transition from NP of $N_{-}
$ to bistable NP$_{bi}$($N_{-}$, $N_{+}$) takes place at the critical point $%
g_{c+}=\sqrt{10/3}$ [determined from Eq. (23)] and then the transition from
bistable NP$_{bi}$($N_{-}$, $N_{+}$) to the NP of $N_{+}$ follows at the
critical point $g_{c-}=4\sqrt{2/3}$ [obtained from Eq. (22)]. Even though
the order-parameter is zero $n_{p\mp }=0$ in both sides of critical point,
while ground state structure changes.

\textbf{\ }
\begin{figure}[t]
\includegraphics[width=2.0in]{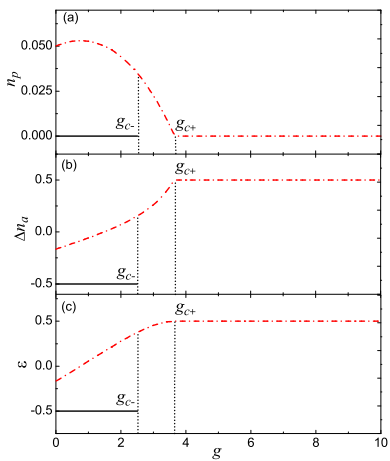}
\caption{(Color online) $n_{p}$ (a), $\Delta n_{a}$ (b), and $\protect%
\varepsilon $ (c) curves in red detuning for $U=$ $-20$. The QPT from NP of $%
N_{+}$ to the SP of $S_{+}$ is in the inverted direction.}
\label{fig.5}
\end{figure}

\begin{figure}[t]
\includegraphics[width=2.0in]{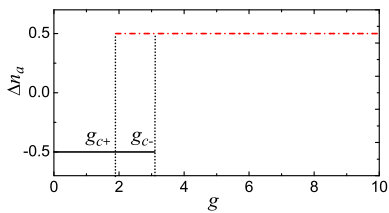}
\caption{(Color online) The QPT among NPs of different types showing by the
variation of atomic population imbalance $\Delta n_{a}$ for $U=$ $-14.$}
\label{Fig.6}
\end{figure}

\section{Conclusion and discussion}

The system of BEC in an optical cavity provides a marvelous model for reveal
of QED theory. Although having investigated extensively the new phenomena
emerge surprisingly, among which the population inversion with full
occupation of excited state and inverted QPT are most exciting. The bistable
MMQSs can be observed experimentally by tuning the frequency of pump laser
and the atom-photon interaction strength, which play central roles in the
observations. The SP of $S_{+}$ for the inverted pseudospin state ($\Uparrow
$) exists only when the effective frequency becomes negative $\omega <0$
along with the nonzero atom-photon interaction. It should be noticed that
when the interaction tends to zero $U$ $\rightarrow 0$ the solutions for
normal state ($\Downarrow $) reduce exactly to those of the standard DM. We
also remark that the SCS variational method has advantage in the theoretical
investigation of macroscopic quantum properties of the atom-ensemble and
cavity-field system, since it results in a one-parameter variational
energy-function to achieve rigorously the analytic solutions of the MMQS.
More importantly the inverted pseudospin state ($\Uparrow $) comes into the
formulation in a natural way giving rise to the bistable phases, which
predicted in the present paper are in agreement with the semiclassical
dynamics of nonequilibrium DM \cite{BMS12}. They were also realized recently
by two of us (N.L and J.Q.L) in a time-driving nonequilibrium model, where
multiple local minima of energy functional are found by the numerical
simulation \cite{NIJL13}.

\section*{ACKNOWLEDGEMENTS} This work is supported by the National
Natural Science Foundation of China, under Grant No. 11275118 and
the Research Training Program for Undergraduates of Shanxi
University (Grant No 2014012174).\ \

\end{document}